\documentclass[review,a4paper,3p,12pt]{elsarticle}

\makeatletter
\def\ps@pprintTitle{%
	\let\@oddhead\@empty
	\let\@evenhead\@empty
	\let\@oddfoot\@empty
	\let\@evenfoot\@oddfoot
}
\makeatother

\usepackage[T1]{fontenc}
\usepackage[utf8]{inputenc}
\usepackage{ams math,amssymb}
\usepackage{algorithm} 
\usepackage{algpseudocode} 
\usepackage[colorlinks=true,citecolor=ForestGreen,linkcolor=Red,urlcolor=Blue,hypertexnames=false]{hyperref}
\usepackage{chngcntr}
\usepackage{subfigure}
\usepackage{graphicx}
\usepackage{ulem}

\usepackage{natbib}
\usepackage[dvipsnames]{xcolor}
\usepackage{hyperref}
\usepackage{booktabs}
\usepackage{color}
\usepackage{setspace} 

\counterwithout{figure}{section}

\begin{document}

\begin{frontmatter}
\title{A Theoretical Model of False Information Control}
\author[1]{Yu Zhang$^*$}
% \ead{zhangyu@ifi.uzh.ch}

% \author[2]{Yafei Li}
\author[3]{Fanyuan Meng}
\author[1]{Vallarano Nicolò}
\author[1]{Claudio J. Tessone}

\cortext[cor1]{Corresponding email: zhangyu@ifi.uzh.ch}

\affiliation[1]{organization={Blockchain and Distributed Ledger Technologies, Institute of Informatics, University of Zürich},
addressline={Andreasstrasse 15},
postcode={8050},
city={Zürich},
country={Switzerland}}
% \affiliation[2]{organization={Agroecology and Environment, Agroscope},
% addressline={Reckenholzstrasse 191},
% postcode={8046},
% city={Zürich},
% country={Switzerland}}

\affiliation[3]{organization={Research Center for Complexity Sciences, Hangzhou Normal University},
postcode={311121},
city={Hangzhou, Zhejiang},
country={China}}

% \author{
% Yu Zhang
% \thanks{zhangyu@ifi.uzh.ch}
% \thanks{Blockchain and Distributed Ledger Technologies Group, Department of Informatics, University of Zurich}
% ,
% Fanyuan Meng 
%   \thanks{Research Center for Complexity Sciences, Hangzhou Normal University}
%  ,
% Vallarano Nicolò
%   \protect\footnotemark[2]
% ,
% Claudio J. Tessone
% \protect\footnotemark[2]
% \thanks{UZH Blockchain Center, University of Zurich}
% }

\begin{abstract}
When considering a specific event, news that accurately reflects the ground truth is deemed as real information, while news that deviates from the ground truth is classified as false information. False information often spreads fast due to its novel and attention-grabbing content, which poses a threat to our society. By extending the Susceptible-Infected (SI) model, our research offers analytical decision boundaries that enable effective interventions to get desirable results, even when intermediate functions cannot be analytically solved. These analytical results may provide valuable insights for policymakers in false information control. When assessing intervention costs using the model, the results indicate that the sooner we intervene, the lower the overall intervention cost tends to be.
\end{abstract}


\end{frontmatter}
\section{Introduction}
Online social networks have revolutionized the way we connect with friends and family, as well as the way we share information with people around the world \cite{heidemann2012online}. However, the ease and speed of information dissemination on these networks have also made them a breeding ground for misinformation, rumor, and disinformation \cite{tambuscio2015fact,zhaofake}. 
The World Economic Forum's Global Risks Report (2023) \cite{wef} has identified digital misinformation and disinformation as one of the most severe risks to social cohesion in both the short term (2 years) and the long term (10 years) scales. Apart from eroding social cohesion, false information can also cause a host of other detrimental effects such as lower vaccination rates due to false information about adverse reactions to vaccines, election disruptions, and damage to government credibility \cite{shu2020mining}.

The variety in spreading intention, channels, validity, and sources makes it difficult to understand their spreading online \cite{vosoughi2018spread}. While both are false in terms of their contents, misinformation is shared inadvertently without the intention to deceive, and disinformation is shared intentionally to deceive \cite{stahl2006difference,aimeur2023fake, tudjman2003information, petratos2021misinformation, wardle2017information}. European Democracy Action Plan took both disinformation and misinformation as fake news, which is similar to the classification in some papers \cite{aimeur2023fake}. \cite{suntwal2020does} even classified hoaxes, rumors, parodies, incorrect editorials, and incorrect facts as fake news, but others have their definition \cite{bondielli2019survey}. Though the intent and authenticity of the news content are the most common key features to classify fake news \cite{aimeur2023fake}, there is still no consensus about whether these items should be taken as fake news or not.

The first definition of fake news was provided by \cite{allcott2017social}, encompassing those news that are intentionally and verifiably false. However, \cite{sharma2019combating} defined fake news as a news article or message carrying false information regardless of the means and motives behind it. The definition of fake news in \cite{bondielli2019survey} is similar to that in \cite{allcott2017social}, but \cite{bondielli2019survey} also distinguished fake news and rumor, which is against the classification in European Democracy Action Plan and other research works \cite{aimeur2023fake}.
In \cite{bondielli2019survey}, a rumor is defined as an unverified claim made by users on social platforms. This claim may turn out to be true or false and is not correlated to intention. In \cite{bondielli2019survey}, both fake news and rumors are classified as misinformation. \cite{gelfert2018fake} thinks that fake news is not only intentionally and verifiably false, but also must be designed systematically. 
The term ‘fake news’ is often especially regarded as problematic \cite{van2022misinformation} and has become a politicized word in itself \cite{van2020you}. Sometimes, the term 'fake news' even started to mean the news that politicians don't like \cite{nakov2020can}.

As \cite{mazur9definition} said, there is still no consistent definition for it \cite{nakov2020can,gelfert2018fake}, not only because of the complex landscape of sub-categories (disinformation, rumor, hoax, rumor, satire, conspiracy theories) but also because of the law issues behind, especially regarding the right of speech freedom. 

Amidst the huge number of fake news (hereby referring to all possible subcategories implied by the term), we chose to focus the scope of the present work on a very specific case of news spreading, namely the contemporaneous spreading of two opposite news.
 There exist some situations where not believing the truth is synonymous with believing the opposite content. For example, in the case of COVID-19, not believing the effectiveness of vaccines implies believing that they are ineffective \cite{guay2022think}. To study these competitive and opposite news spreading, research using epidemic-based models, like Susceptible-Infected (SI) or Susceptible-Infected-Recovered (SIR) was developed \cite{he2015modeling, zhao2011rumor,zhu2020delay, maleki2021using}. For example, in \cite{he2015modeling}, authors take into consideration not only the spreading of rumor but also the spreading of immunization to the rumor at the same time using the SIR model. In \cite{acerbi2022research},  the authors simulated the effectiveness of intervention between promoting real information and fighting misinformation. Although they didn't use an SI- or SIR-based model, the model setting is very similar to that in our paper.

In a similar fashion to \cite{he2015modeling, zhao2011rumor,zhu2020delay, maleki2021using}, distinguishing and defining the concepts of fake news, disinformation, misinformation, rumor, et al., is not the focus of the present work. 
In the model we present, news reflecting the ground truth of an event is classified as real information, otherwise, false information. By this definition, the real information and false information in this paper are only content-based and are not related to their spreading intent, motivation, source, etc al.. In the example of the effectiveness of the COVID-19 vaccine, the real information would be that 'COVID-19 vaccine is effective' and the false information would be that 'COVID-19 vaccine is effective'.

For any event, the relationship between its real information and false information is the same as the relationship between the original proposition and its corresponding converse proposition. For example, the specific event is 'A caused B'. Then, the real information is 'A caused B', and the corresponding false information is 'A didn't cause B'. Similarly, if the original proposition is 'A caused B', then its corresponding converse proposition is 'A didn't cause B'. In our model, we will only take into consideration the spreading of specific news and its opposite news. In reality, it is not always easy to distinguish between real and false information from a common sense perspective, and even to conduct fact-checking \cite{mahir2019detecting}. However, as we mentioned above, distinguishing the real and false information is not the focus of this paper. Therefore, for the sake of description, in our model, we refer to one of them as false information and the opposite one as real information. 

False information disseminates very fast through online social networks \cite{sharma2019combating, qian2018neural} and some papers find that false information spreads faster than real information \cite{buechel2023misinformation,bovet2019influence}, sometimes can be even six times faster \cite{vosoughi2018spread} because the content of false information is usually more novel than the real information \cite{buechel2023misinformation,bovet2019influence,vosoughi2018spread}. 
Without external intervention, a large number of people would believe false information, which is damaging to society. External intervention can be an endeavor to stop the spreading of false information or to speed up the spreading of real information beyond the spreading of false information \cite{acerbi2022research}. However, the current trend in research predominantly revolves around combatting the proliferation of misinformation through methods such as fact-checking, misinformation inoculation, pre-bunking, and debunking and there is little research that focuses on speeding up the spreading of real information \cite{acerbi2022research}. In the present work, we assume that external intervention can only accelerate the dissemination of real information to individuals who have not yet been exposed to any news, encompassing both false and real information, and such intervention can't compel individuals to alter their attitudes once they have formed the belief that the news they have encountered is true in short term. Because both real and false information spreads in the same network, accelerating the spreading of real information also means slowing down the dissemination of false information. Therefore, we only focus on the speeding up of real information spreading. Given this situation, the challenge lies in determining the appropriate time and strength of intervention to accelerate the number of individuals who trust the real information so that desirable results can be obtained. Considering this problem, we explore the decision boundary by extending the Susceptible-Infected (SI) framework which is used to model the spreading of viruses among humans, animals, and plants \cite{gilligan2008sustainable}. The decision boundary here is defined as a function of the intervention strength with the disparity between the percentage of individuals who believe false information and those who believe real information.

Extensive research has been conducted on the spreading of (mis)information. A primary focus of these studies is controlling the spreading of false information by e.g. identifying the influential nodes in the network \cite{chaudhury2012spread, alassad2019finding}, and recognizing the false information by their properties \cite{shu2019studying, zhou2019network, ruchansky2017csi}. While it was found that the intervention methods in the two directions were full of challenges. For example, \cite{budak2011limiting} proved that minimizing the spreading of misinformation in social networks is an NP-hard problem. Other methods that were applied to fight against misinformation or fake news include fact-checking, misinformation inoculation, pre-bunking, and debunking \cite{traberg2022psychological,roozenbeek2019fake,lewandowsky2021countering}. One big difference between fact-checking, debunking, and those like pre-bunking and inoculation is the timing. Inoculation, which includes threat and refutational preemption, is more complicated than pre-bunking. However, because of the large number of falsehoods online every day, it is not realistic to make pre-bunking/debunking/fact-checking/inoculating for each piece of falsehood. \cite{traberg2022psychological}.

Theoretical models of news spreading such as the Daley-Kendal model were constructed in \cite{daley1964epidemics} and developed in \cite{maki1973mathematical}. Epidemic models have also been applied to research the diffusion of information \cite{he2015modeling, zhao2011rumor,zhu2020delay, maleki2021using, wang2013rumor, liao2014can,jin2013epidemiological} and epidemiological models can provide a classical approach to study how information diffuses \cite{jin2013epidemiological}. As \cite{fan2020novel} stated, "the spreading of fake news and rumors can be seen as ‘infection of the mind’". Therefore, the SI-related models are a suitable choice for studying the spreading of news \cite{de2015exploiting, abdullah2011epidemic}. However, these researches primarily focus on the spreading of false information only. To the best of our knowledge, there is still a dearth of research that simultaneously investigates the spreading of both false information and information, particularly in exploring the decision boundary of intervention of false information. Our research provides an analytical solution for the decision boundary, offering insights into the time and strength of intervention, which enables us to achieve desired results even in scenarios where the intermediary functions can not be analytically resolved. After conducting numerical simulations using both the Erdős–Rényi (ER) network and Barabási–Albert (BA) network, we find that though the shape of decision boundaries is similar to the theoretical one, the corresponding values of ER and BA are not equal to the theoretical value, which indicates that the network structure does matter. Here the theoretical intervention model we constructed can also be applied in other fields, like viral marketing economics \cite{rodrigues2015viral}.
In viral marketing economics, the message about the product is like a virus that spreads among customers by word-of-mouth and this model can be used to simulate and assess the effectiveness of promoting strategy when there are competitors in the same market. 

The remainder of this paper is organized as follows. In section 2, we present conceptual models. In section 3, we formulate mathematical models with and without intervention and provide an analytical solution for the decision boundary. In section 4, we extend the model to various scenarios and provide corresponding solutions. Section 5 presents the results of numerical simulations on ER and BA networks to compare the experimental boundary with the theoretical boundary. In the final section, we conclude and discuss the strengths and weaknesses of our models, and propose future research directions.

% \section{Model Building Based on SI (Susceptible-Infected) Model}
\section{The model}
Among a population of $n$ agents, real and false information compete. For the simplicity of description in the following section, we define the following three terms, false agent, real agent, and susceptible agent.
\begin{description}
\item[False agent] If an agent believes false information, then it is called a false agent who will spread false information to its susceptible neighbors. Once one agent becomes a false agent, it needs much time to change her/his attitude. However, the news spreads fast on social networks, and it takes a very short time for news to reach everyone on the online social network. Thus, in our model, we assume that once agents become false agents, they will be false agents forever in our model for the sake of simplicity.
\item[Real agent] If an agent consistently believes real information, then it is called a real agent. The real agent will spread real information to their susceptible neighbors. Once an agent attains the status of a real agent, it maintains this category in our model.
\item[Susceptible agent] If an agent is neither a false agent nor a real agent, it is termed a susceptible agent who hasn't heard about or trusted any news yet. Susceptible agents can only become real agents, false agents, or is still susceptible agent after hearing real or false information.
\end{description}

At time step $t=0$, a proportion of susceptible agents receives either real or false information, leading to the formation of initial real agents ($r(t_0)$) or false agents ($f(t_0)$), respectively. This distribution of agents serves as our initial condition.
At the subsequent time step ($t=1,2,\dots$), both real agents and false agents spread their believed news to their respective susceptible neighbors at different rates, i.e. $\beta_r$ and $\beta_f$. The fraction of agents believing real information, false information, and in a susceptible state at time $t$ can be denoted by $r(t)$, $f(t)$, and $s(t)$. Due to the novelty of false information's content, in some cases, the false information may spread faster than real information, i.e. $\beta_f>\beta_r$. For the simplicity of describing the model, we assume this is true now, but we will relax this assumption later. We assume that the competing process will be intervened by some external intervention at the time step $\tau$ when the difference between $f(t)$ and $r(t)$ meets a threshold condition $g$. The dynamic process ceases until all agents become real agents or false agents. 

\section{Results} \label{section3}
\subsection{Before intervention}
Based on the model description, we can articulate the motion of news propagation as follows
\begin{equation}\label{eq:motion}
\left \{
\begin{aligned}
& \frac{df(t)}{dt}=\beta_f \langle k \rangle s(t)f(t),  \\
& \frac{dr(t)}{dt}=\beta_r \langle k \rangle s(t)r(t), \\
& f(t)+r(t)+s(t)=1,
\end{aligned}
\right.
\end{equation}
where $ \langle k \rangle $ is the average degree of the network.
Hence Eq.\ref{eq:motion} can be solved as
\begin{equation} \label{eq:f(t)}
    f(t) = \frac{f(0)}{r(0)^\alpha}r(t)^\alpha.
\end{equation}

where $\alpha=\beta_f/\beta_r$. 

According to Eq.\ref{eq:f(t)}, we could have
\begin{equation}\label{equation3}
    \frac{d \Big(\frac{f(t)}{r(t)}\Big)}{dt}=\frac{f(0)}{r(0)^\alpha}(\alpha-1) \frac{dr(t)}{dt},
\end{equation}

Since we assume the false information spreads faster than real information, i.e. $\beta_f > \beta_r$, we have $\alpha-1>0$. In addition, as the number of agents who adopt real information is increasing, $dr(t)/dt>0$ is realized. Then we must have $d\Big(\frac{f(t)}{r(t)}\Big)/dt>0$ which implies that $g = f(t)-r(t)$ increases with time. 

Because the susceptible agents will become real agents or false agents after they hear about real information or false information, and the real agents and false agents can not convert to susceptible agents in this model, so, there will be only real agents and false agents when the system becomes stable. From the mathematical point of view, a system becomes stable only when $\frac{df(t)}{dt}=0$, $\frac{dr(t)}{dt}=0$, and $\frac{ds(t)}{dt}=0$. The solution is $s(t)=0$, namely, we have $s(\infty)=0$ and $f(\infty)+r(\infty)=1$ in the end. 
Thus from Eq.\ref{eq:f(t)} we could obtain the following equation
\begin{equation}
    f(\infty) = \frac{f(0)}{r(0)^\alpha}{(1-f(\infty))}^\alpha.
\end{equation}
\subsection{After intervention and the decision boundary}
We assume that when the fraction of people receiving false information is higher than that of receiving real information by the threshold $g$, i.e. $f(t)-r(t)=g$, external intervention starts and the intervention intends to speed up the spreading of real information. 

The external intervention methods can be broadcasting the real information to all agents using authoritative radio or TV channels. Because the fact is always that it is not easy for people to change their viewpoints once they believe what they believe is true in a short time and the news spreads very fast on online social networks, thus, we assume that the broadcast will not affect the attitudes of false agents or real agents, but it will solely influence susceptible agents. The corresponding effect of broadcasting is that an extra constant proportion $b$ of susceptible agents ($s(t)$) believe the real information in each further time step. The logic here is similar to \cite{rossman2021network}. Assuming at time $\tau$, the condition of external intervention satisfies, i.e. $f(\tau)-r(\tau)=g$, then we could have $f(\tau) = \frac{f(0)}{r(0)^\alpha}(f(\tau)-g)^\alpha$.
% at time $\tau$, which is also the initial condition of the new propagation model. 

As external intervention starts after time $\tau$ by broadcasting via e.g. TV or radio, an extra constant proportion $b$ of susceptible agents ($s(t)$) will believe the real information in each time step. The model after the intervention is
\begin{equation}\label{eq:new_motion}
\left \{
\begin{aligned}
& \frac{df(t)}{dt}=\beta_f \langle k \rangle s(t)f(t),  \\
& \frac{dr(t)}{dt}=\beta_r \langle k \rangle s(t)r(t)+b s(t), \\
& f(t)+r(t)+s(t)=1,\\
& f(\tau)-r(\tau)=g.
\end{aligned}
\right.
\end{equation}
where $b$ ($0\le b\le 1$) is the strength of intervention, $bs(t)$ measures the effect of broadcasting (intervention).
Then Eq.\ref{eq:new_motion} can be solved as
\begin{equation}
    f(t)=C_1 \Big(r(t)+\frac{b}{\beta_r\langle k \rangle}\Big)^{\alpha},
\end{equation}
where $C_1=\frac{f(0)}{r(0)^\alpha}\Big[\frac{ r(\tau)}{r(\tau)+\frac{b}{\beta_r \langle k \rangle}}\Big]^\alpha$ and $\alpha=\beta_f/\beta_r$.

When the system is stable ($s(t)=0$), we assume that $f(\infty)=r(\infty)=\frac{1}{2}$ is realized.  Then we have

\begin{equation} \label{eq:extreme}
\left \{
\begin{aligned}
& f(\infty)=\frac{f(0)}{r(0)^\alpha}\Big[\frac{ r(\tau)}{r(\tau)+\frac{b}{\beta_r\langle k \rangle}}\Big]^\alpha \Big(r(\infty)+\frac{b}{\beta_r\langle k \rangle}\Big)^{\alpha},  \\
& r(\infty)=\frac{1}{2}, \\
& f(\infty)=\frac{1}{2}.
\end{aligned}
\right.
\end{equation}

From Eq.\ref{eq:extreme}, we get the analytical solution of $r(\tau)$ as
\begin{equation}
    r(\tau)=\frac{b\frac{(\frac{1}{2})^{\frac{1}{\alpha}}}{\frac{1}{2}+\frac{b}{\beta_r \langle k \rangle}}}{\beta_r \langle k \rangle (\frac{f(0)^{\frac{1}{\alpha}}}{r(0)}-\frac{(\frac{1}{2})^{\frac{1}{\alpha}}}{\frac{1}{2}+\frac{b}{\beta_r \langle k \rangle}})}.
\end{equation}
Since $\frac{f(0)}{r(0)^\alpha}r(\tau)^\alpha-r(\tau)=g$, then we get the relationship between $g$ and $b$ ($g=G(b)$) as

\begin{equation}
   g=\frac{f(0)}{r(0)^\alpha} \Big[\frac{b\frac{(\frac{1}{2})^{\frac{1}{\alpha}}}{\frac{1}{2}+\frac{b}{\beta_r \langle k \rangle}}}{\beta_r \langle k \rangle (\frac{f(0)^{\frac{1}{\alpha}}}{r(0)}-\frac{(\frac{1}{2})^{\frac{1}{\alpha}}}{\frac{1}{2}+\frac{b}{\beta_r \langle k \rangle}})}\Big]^\alpha-\frac{b\frac{(\frac{1}{2})^{\frac{1}{\alpha}}}{\frac{1}{2}+\frac{b}{\beta_r \langle k \rangle}}}{\beta_r \langle k \rangle (\frac{f(0)^{\frac{1}{\alpha}}}{r(0)}-\frac{(\frac{1}{2})^{\frac{1}{\alpha}}}{\frac{1}{2}+\frac{b}{\beta_r \langle k \rangle}})},
\end{equation}
where $0\le b\le 1$. We can find that the decision boundary is also a function of the average degree of the network $ \langle k \rangle $. If $b=0$, then the model will reduce to the case without intervention.

\begin{figure}[h]
    \centering
    \includegraphics[width=0.7
    \linewidth]{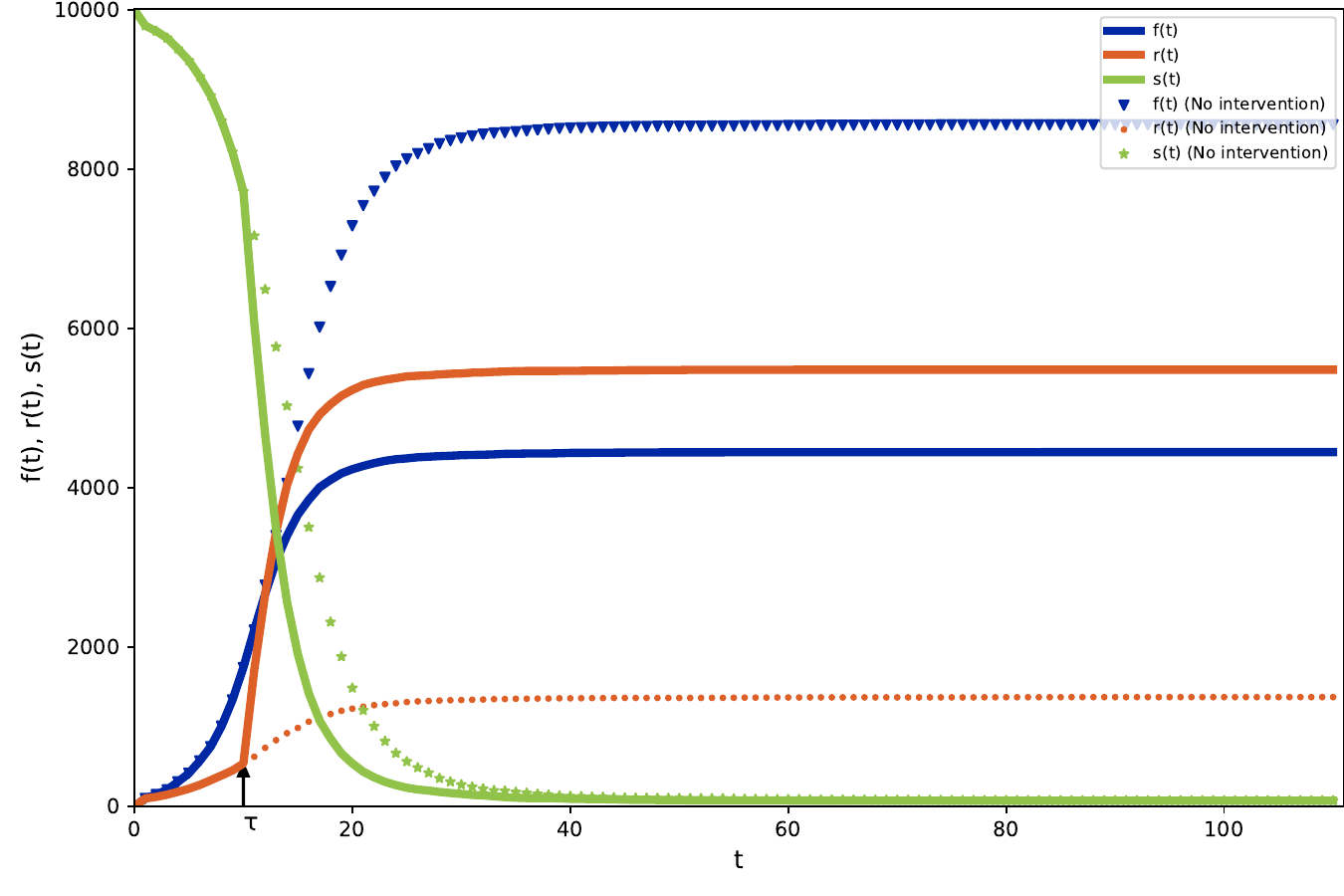}
    \caption{Numerical simulation of news spreading with and without intervention. The three dotted lines in the figure denote the number of a real agent, false agent, and susceptible agent, respectively, in the case that there is no external intervention; the three solid lines denote the evolving number of corresponding agents in the case that we intervene with $b$ at $\tau$, namely the condition $\frac{f(0)}{r(0)^\alpha}-r(\tau)=g$ is satisfied at $\tau$. } 
    \label{s1}
\end{figure}

Here, we give an example to illustrate what will happen before and after we intervene on social networks. The experiment is done on a random network with ten thousand and one agents whose average degree is 5. Both $f(0)$ and $r(0)$ are set as 0.01, $\beta_f$ and $\beta_r$ are set as 0.1 and 0.05, respectively, so that $\beta_f>\beta_r$. If there is no intervention, there is a very large probability that the number of false agents is much larger than the number of real agents when $s(t)=0$, as shown by the blue dotted line and orange dotted line in Fig.\ref{s1}. Now, when $f(t)-r(t)=0.1$ at time $\tau$, we set $b=0.2$ so that we speed up the spreading of real information after $\tau$. As shown in Fig.\ref{s1}, the solid lines overlap with the dotted lines before $\tau$, after we intervene at $\tau$, the orange solid line is more upright than the blue solid line which means that the real information spread faster than the false information. When $s(t)=0$, the number of real agents is larger than the number of false agents, which means we intervene successfully. Because this is a probability problem, it is still probable in reality that the number of real agents is larger than the number of false agents even if there is no intervention when $s(t)=0$. However, the value of this probability may be very small. The action of intervention will increase this probability. 

%\textcolor{red}{We do not have any description about figures 1-6.}

%\section{General Case}
In the general case, we assume that the goal of the external intervention is that $r(\infty)=P (P \ge r(0))$, $f(\infty)=(1-P)$, then we have
\begin{equation}\label{eq:general_case}
\left \{
\begin{aligned}
& f(\infty)=\frac{f(0)}{r(0)^\alpha}\Big[\frac{ r(\tau)}{r(\tau)+\frac{b}{\beta_r \langle k \rangle}}\Big]^\alpha \Big(r(\infty)+\frac{b}{\beta_r\langle k \rangle}\Big)^{\alpha},  \\
& r(\infty)=P, \\
& f(\infty)=1-P.
\end{aligned}
\right.
\end{equation}

By solving Eq.\ref{eq:general_case}, we get 
\begin{equation}
    r(\tau)=\frac{bm}{\beta_r \langle k \rangle (1-m)},
\end{equation}
where $m=\frac{r(0)}{f(0)^{\frac{1}{\alpha}}} \frac{(1-P)^{\frac{1}{\alpha}}}{P+\frac{b}{\beta_r \langle k \rangle}}$. Then the relationship between $g$, $b$ and $P$ ($g=G(b,P)$) is

\begin{equation}
\label{eq11}
    g=\frac{f(0)}{r(0)^\alpha} \Big[\frac{b m}{\beta_r \langle k \rangle (1-m)}\Big]^\alpha-\frac{b m}{\beta_r \langle k \rangle (1-m)}.
\end{equation}
where $0 \le b \le 1$ and
$m=\frac{r(0)}{f(0)^{\frac{1}{\alpha}}} \frac{(1-P)^{\frac{1}{\alpha}}}{P+\frac{b}{\beta_r \langle k \rangle}}$.
% where $\alpha'$ is a function of $b$.

\begin{figure}[h]
    \centering
    \includegraphics[width=0.7\textwidth]{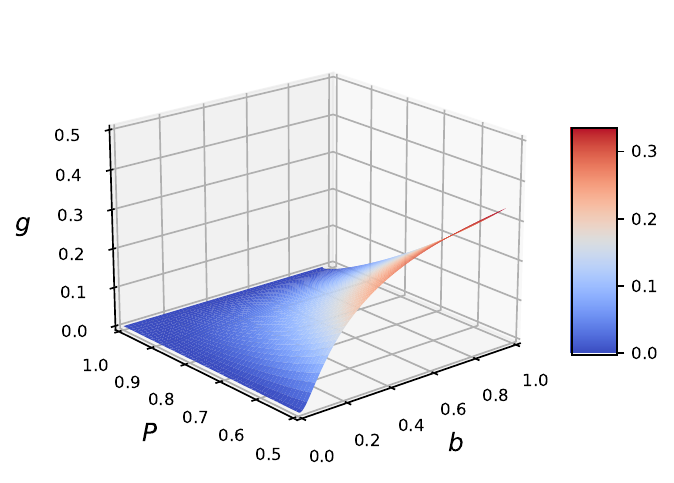}
    \caption{The theoretical decision boundary $g=G(b,P)$. The initial conditions are $r(0)=f(0)=0.01$, $\beta_f=0.1$, $\beta_r=0.05$, $\langle k\rangle=5$.}
    \label{gpb}
\end{figure}

We now discuss the relationship among $g$, $b$, $P$ and $\langle k\rangle$ concisely based on the analytical results we got. It is clear that $b$ will increase with $P$. This relationship makes sense and is shown in Fig.\ref{gpb} in the numerical simulation section. 

Now, we explain why $b$ also increases with $g$. Based on equation \ref{equation3}, since we assume that the false information spreads faster than real information, i.e. $\beta_f > \beta_r$, we have $\alpha-1>0$. Then we  have $d\Big(\frac{f(t)}{r(t)}\Big)/dt>0$ which implies that $g = f(t)-r(t)$ increases with time. 
We also know that $r(t)$, $f(t)$  increase monopoly with time. So, larger $g$ means that the we intervene later,   $s(t)$ is smaller, and importantly more new susceptible agents begin to believe fake news in each time unit but not real news. To get more share of $s(t)$, the intervention strength $b$ has to be larger. 

\begin{figure}[h]
    \centering
    \includegraphics[width=0.45\textwidth]
    {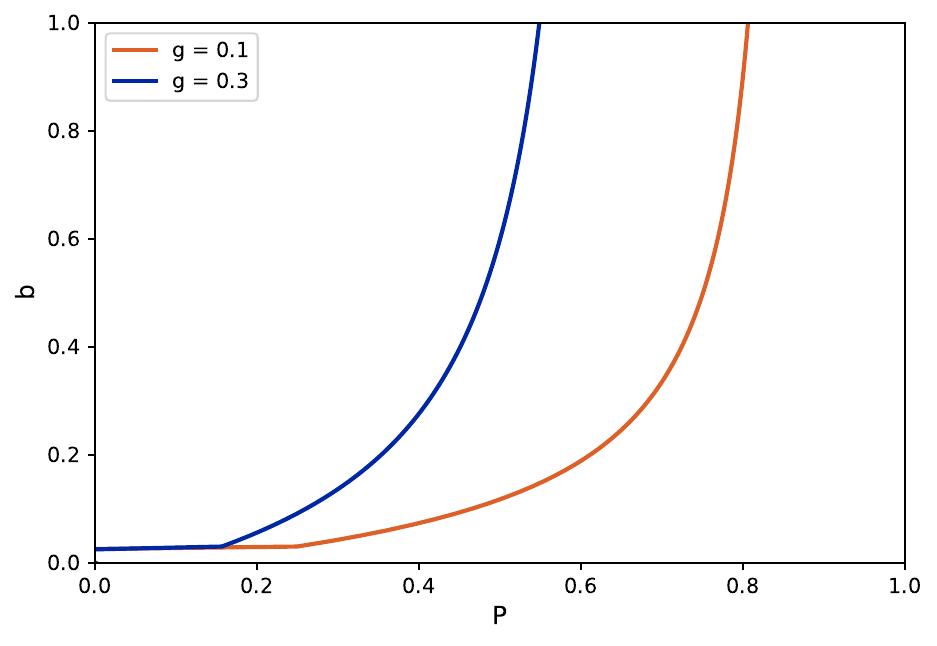}
    \includegraphics[width=0.45\textwidth]{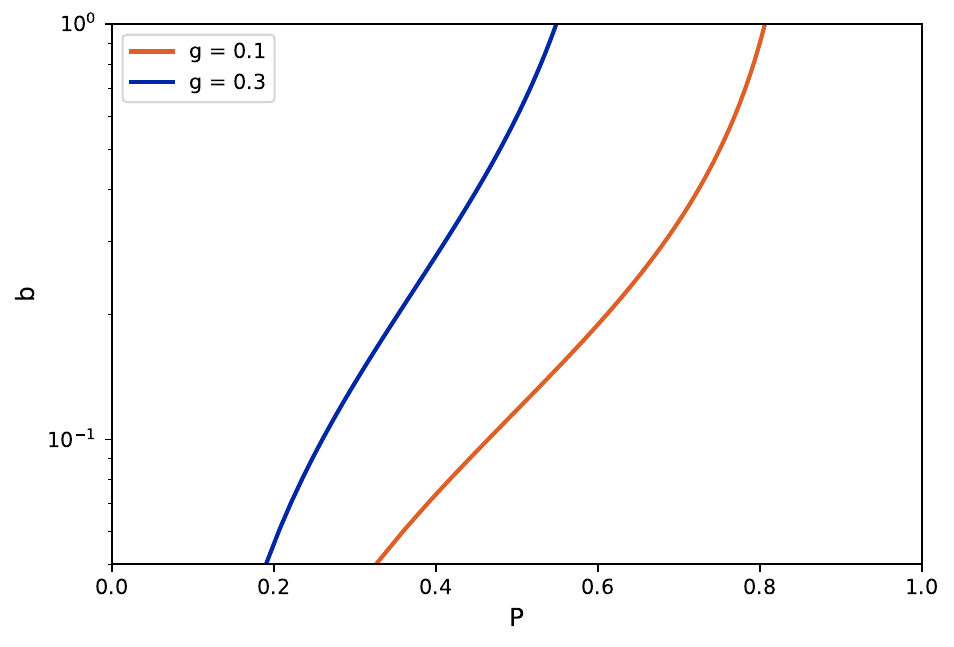}
    \caption{$b$ increases sharply with $P$ for given $g$. The y-axis of the right figure is the log scale. We can find $b$ almost increases exponentially with $P$ for a given $g$. For a larger $P$, $b$ may even increase super-exponentially.}
    \label{inter_strength}
\end{figure}

From equation \ref{eq11}, the intervention strength $b$ is a function of $P$ for given $g$. For the sake of clarity, we plot Fig.\ref{inter_strength} to reflect the relationship between $P$ and $b$ and we find that $b$ increases very fast with $P$. Actually, $b$  increases almost exponentially with $P$ for a given $g$ or even super-exponentially when $P$ is larger, which means that policymakers may need more resources for intervention each time. 

\begin{figure}[h]
    \centering
    \includegraphics[width=0.45\textwidth]{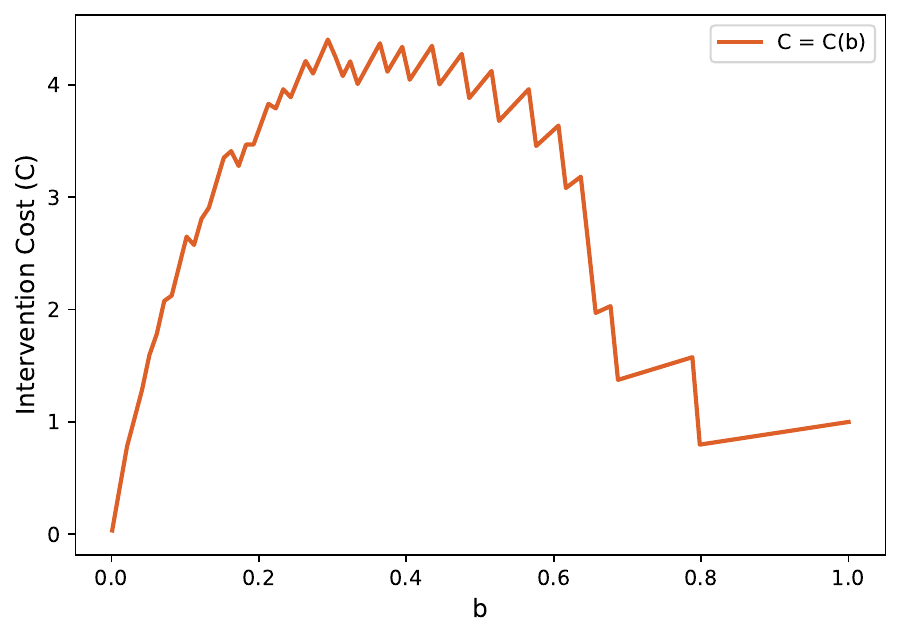}
    \caption{Total intervention cost $C$ change with the intervention strength $b$.}
    \label{inter_cost}
\end{figure}

If we see $b$ as the cost in one time of intervention, the intervention duration also matters. If the intervention strength $b$ is larger, then the duration of the intervention may be shorter, so, what is the relationship between the total cost ($C$) and $P$? It is hard to calculate $C$ analytically. But we can estimate the time $\tau$ when the intervention starts and the ending time $T$ when the percentage of susceptible agents ($S$) is below, like 1\%, after intervening based on our model. Then we can take the number of iteration steps from $\tau$ to $T$ as the duration of intervention.  The total intervention cost can be estimated by $b$ times the number of iteration steps. The result is shown in Fig. \ref{inter_cost}. This method may be not very accurate, but can be an estimation. We notice that the total cost increases first with $b$ then decreases with $b$ after some point. This indicates that if the policymaker what to intervene with a limited budget, then they can either intervene earlier with lower intervention strength $b$ but a longer duration of intervention, or intervene later with higher intervention strength $b$, but a shorter duration of intervention. 

In the cost analysis above, we took the intervention strength $b$ as the cost each time. But in reality, the intervention cost each time may be some monopoly increasing function of the intervention strength $b$. If the function is convex, then the total intervention cost $C$ may increase monopoly with the intervention strength $b$.
However, in both cases, we can find that the sooner the policymaker intervenes, the lower the total intervention cost $C$ should be.

\begin{figure}[h]
    \centering
    \includegraphics[width=0.7\textwidth]{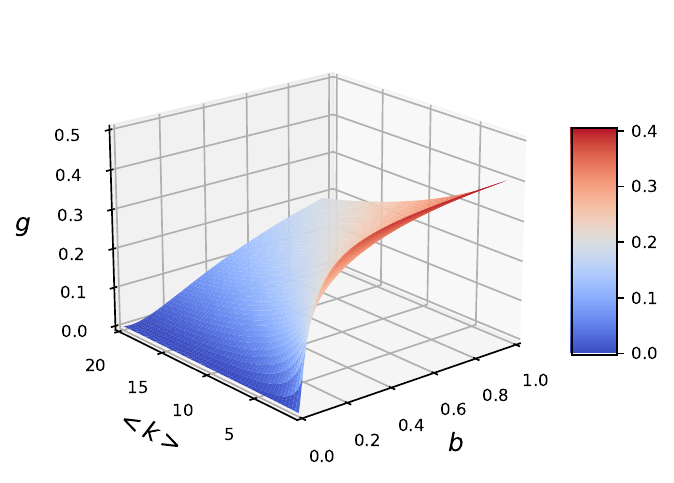}
    \caption{The theoretical decision boundary $g=G(b,\langle k\rangle)$. The initial conditions are $r(0)=f(0)=0.01$, $\beta_f=0.1$, $\beta_r=0.05$, $P=\frac{1}{2}$.}
    \label{gkb}
\end{figure}

It is worth noting the relationship between $b$ and the average degree of social network $\langle k\rangle$. As shown in Fig.\ref{gkb}, if $g$ is fixed, then $b$ increases with $\langle k\rangle$. Thus, the denser the social network is, the stronger external intervention we need. It means that it is more difficult to control the spreading of false information in dense networks than that in sparse networks.

\subsection{General scenario}
When building the model, we assume that the false information may spread faster than real information, i.e. $\beta_f>\beta_r$. There are still many cases where real information spreads faster than false information, but policymakers still want more people to believe the real information, like the information about the effectiveness of the COVID-19 vaccine. 

Under this assumption that real information spreads faster than false information, the model is almost the same as that above. Now, we have. $\beta_f < \beta_r$, and $\alpha-1<0$. In addition, because of  $dr(t)/dt>0$, then we  have $d\Big(\frac{f(t)}{r(t)}\Big)/dt=\frac{f(0)}{r(0)^\alpha}(\alpha-1) \frac{dr(t)}{dt}<0$ which implies that $f(t)-r(t)$ decrease with time.  In this case, the model before intervention and after intervention is also the same as the case above and we can get similar results.

\newpage
\section{Robust checking: other modelling scenarios and their decision boundaries}

In section \ref{section3}, we built a mean-field model to simulate the spreading of real information and misinformation. By assuming that the outer intervention can promote extra $bs(t)$ susceptible agents to believe real information. Based on this setup, we got the analytical decision boundaries that enable effective interventions to get desirable results. We found that this kind of decision boundary could also be gotten analytically in several other cases and the shapes of their decision boundaries were also very similar. 

In the following parts, we listed these different intervention models and their corresponding decision boundaries.

\subsection{Case 1}
What if the real information is broadcast only one time? In this case, we can assume that after the broadcasting, the model is still the same as before. The only changes in this situation are the number of real agents ($r(t)$) increases by another extra $b s(\tau)$ only at time $\tau$ when $f(\tau)-r(\tau)=g$, where $b$ is the intervention strength and $s(\tau)$ is the number of the susceptible agent. After $\tau$, both the false information and real information spread in the same manner as before. The model is

\begin{equation}\label{case1}
\begin{cases}
& \frac{df(t)}{dt}=\beta_f \langle k \rangle s(t)f(t),  \\
& \frac{dr(t)}{dt}=\beta_r  \langle k \rangle s(t)r(t), \\
& f(t)+r(t)+s(t)=1,\\
& f(\tau)-r(\tau)=g,\\
& r(\tau)_{new}=r(\tau)+b s(\tau),\\
& s(\tau)_{new}=(1-b) s(\tau).
\end{cases}
\end{equation}

The new initial condition of the model after intervention is $r(\tau)_{new}$, $f(\tau)$ and $s(\tau)_{new}$. If we assume that $r(\infty)=P (P \ge r(0))$, $f(\infty)=(1-P)$ when the system is stable, then we can get $r(\tau)=\frac{b(1-g)}{1-m(1-2b)}$, where $m=\Big(\frac{(1-P) r(0)^\alpha}{P^\alpha f(0)}\Big)^{\frac{1}{\alpha}}$ and $\alpha=\frac{\beta_f}{\beta_r}$. The decision boundary is
\begin{equation}
    g=\frac{f(0)}{r(0)^\alpha} \Big[\frac{b(1-g)}{1-m(1-2b)}\Big]^\alpha-\frac{b(1-g)}{1-m(1-2b)}.
\end{equation} 

The shape of the decision boundary is like that in section \ref{section3}. However, because the policymakers only intervene once in this case, so, the intervention strength $b$ can be taken as the total intervention cost. Just like that in Figure \ref{inter_strength}, the total intervention cost now increases with $b$ exponentially, which means that the later the regulator intervenes, the larger the cost is.

\subsection{Case 2}

In the above models, we always think that the broadcasting will not affect the spreading rate $\beta_r$. However, broadcasting can also increase the credibility of real information to those real agents, then these real agents may also strengthen their endeavor in spreading real information. For example, if real agents spend 1 hour to persuade their neighbors to believe real information before intervention. After broadcasting, they may spend 2 or more hours, which will increase the spreading rate of real information.
As external intervention starts after time $\tau$, the spreading rate of real information will increase by some value $b$. Then,
the model is
\begin{equation}\label{eq:case2}
\left \{
\begin{aligned}
& \frac{df(t)}{dt}=\beta_f \langle k \rangle s(t)f(t),  \\
& \frac{dr(t)}{dt}=(\beta_r+b) \langle k \rangle s(t)r(t), \\
& f(t)+r(t)+s(t)=1,\\
& f(\tau)-r(\tau)=g.
\end{aligned}
\right.
\end{equation}

When the system is stable ($s(t)=0$), we assume that the goal of the external intervention is that $r(\infty)=P (P \ge r(0))$, $f(\infty)=(1-P)$, then we have
\begin{equation}\label{eq16}
\left \{
\begin{aligned}
& f(\infty)=\frac{f(0)}{r(0)^\alpha} r(\tau)^{\alpha-\alpha'} r(\infty)^{\alpha'},  \\
& r(\infty)=P, \\
& f(\infty)=1-P.
\end{aligned}
\right.
\end{equation}
where $\alpha'=\frac{\beta_f}{\beta_r+b}$ and $0 \le b \le 1-\beta_r$. 

By solving Eq.\ref{eq16}, we get 
    $r(\tau)=\frac{f(0)^{\frac{1}{\alpha'-\alpha}}}{r(0)^{\frac{\alpha}{\alpha'-\alpha}}} (\frac{1-P}{P^{\alpha'}})^{\frac{1}{\alpha-\alpha'}}$.
Then the relationship between $g$, $b$ and $P$ ($g=G(b,P)$) is
\begin{equation}
\label{eq17}
    g=\frac{f(0)^{\frac{\alpha'}{\alpha'-\alpha}}}{r(0)^{\frac{\alpha \alpha'}{\alpha'-\alpha}}} \Big(\frac{1-P}{P^{\alpha'}}\Big)^{\frac{\alpha}{\alpha-\alpha'}} -\frac{f(0)^{\frac{1}{\alpha'-\alpha}}}{r(0)^{\frac{\alpha}{\alpha'-\alpha}}} \Big(\frac{1-P}{P^{\alpha'}}\Big)^{\frac{1}{\alpha-\alpha'}}.
\end{equation}
where $\alpha'=\frac{\beta_f}{\beta_r+b}$.

Under the model setup in this case, both the analytical decision boundary and the estimated total intervention cost (using the same estimation method as that in Section 3) are similar to that (Fig. \ref{s3} and Fig. \ref{inter_cost}) in Section 3.

\subsection{Case 3}
Some researchers may argue that even if the real information is broadcast to everyone via e.g. TV or radio, it is still hard to affect the spreading rate as in case 2 in reality. We know that $\langle k \rangle$ is the average connection of one agent to other agents if we ignore the structures of the social networks. Under this situation, we can argue that the $\langle k \rangle$ of real information can be added by some value $b$. The logic is that people who believe the real information build more connections with other people and share the real information more than before the intervention. The new model is

\begin{equation}\label{case1}
\begin{cases}
& \frac{df(t)}{dt}=\beta_f \langle k \rangle s(t)f(t),  \\
& \frac{dr(t)}{dt}=\beta_r (\langle k \rangle+b) s(t)r(t), \\
& f(t)+r(t)+s(t)=1,\\
& f(\tau)-r(\tau)=g.
\end{cases}
\end{equation}

This equation is very similar to the case 2 model and can be solved analytically using the same process as that in case 2. The problem with this model is that it is hard to simulate it in a network because the network structure needs to change at each step.

\section{Numerical simulation experiment}

We did simulation experiments on both ER and BA networks to check if the theoretical boundary we got in section \ref{section3} is consistent with the real social networks.
ER network is a random model, each edge is linked with a fixed probability. Normally, the degree distribution of the ER network is Poisson, which is not always consistent with real-world observations. However, it is considered a standard baseline for network properties comparison. The BA network is a scale-free model that is based on the preferential attachment mechanism, which can be used to explain many real networks, like the World Wide Web network and the citation network.

\begin{figure}[ht]
    \centering
    \includegraphics[width=0.7\textwidth]{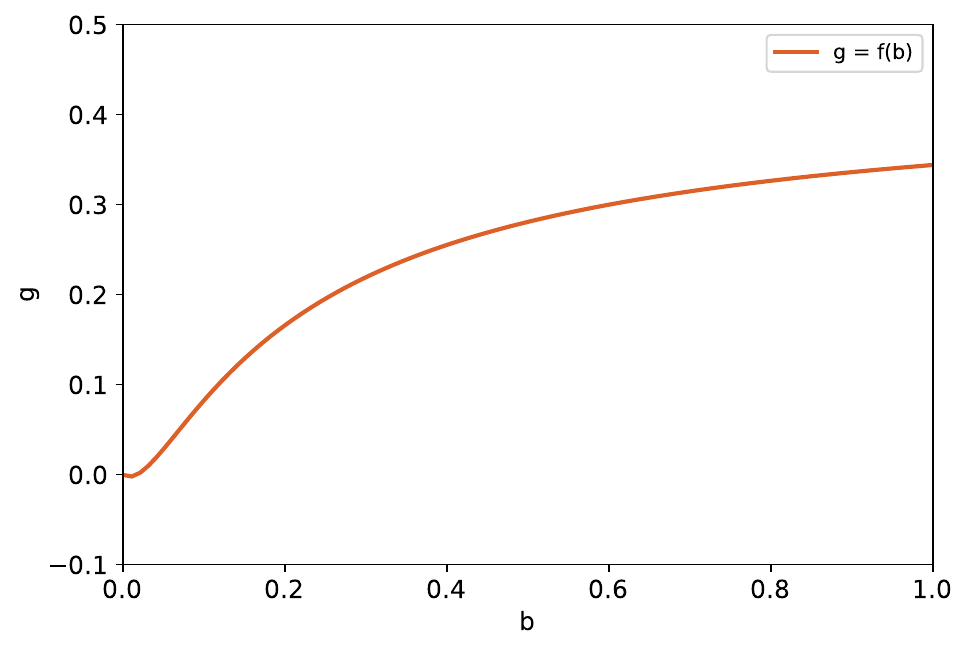}
    \caption{The theoretical decision boundary $g=G(b)$. The initial conditions are $r(0)=f(0)=0.01$, $\beta_f=0.1$, $\beta_r=0.05$}
    \label{s3}
\end{figure}

The total number of agents ($N$) we set for both kinds of networks is ten thousand and one, and the average degree is 5. Both $f(0)$ and $r(0)$ are set as 0.01, $\beta_f$ and $\beta_r$ are set as 0.1 and 0.05, respectively, so that $\beta_f>\beta_r$. The theoretical decision boundary under this setting-up is shown in Fig.\ref{s3}. 

To get the decision boundary (relationship between $g$ and $b$) on both ER and BA networks, we do the simulations according to the following algorithm.
\begin{enumerate}
    \item Under the specific value of $g_{\tau}$ and $b$, we first simulate the spreading of false information and real information based on the original rate $\beta_f$ and $\beta_r$. When $r(\tau)-f(\tau)=g_\tau$,  we increase the spreading rate of real information by $b$, namely $\beta_r=\beta_r+b$ after $\tau$. The result is 1 if the intervention is successful in the end, -1 otherwise.
    \item The experiment is repeated one hundred times under each combination of $b$ and $g$ and we calculate the results on average.
\end{enumerate}

\begin{figure}[h] 
\centering
\includegraphics[width=0.45\linewidth]{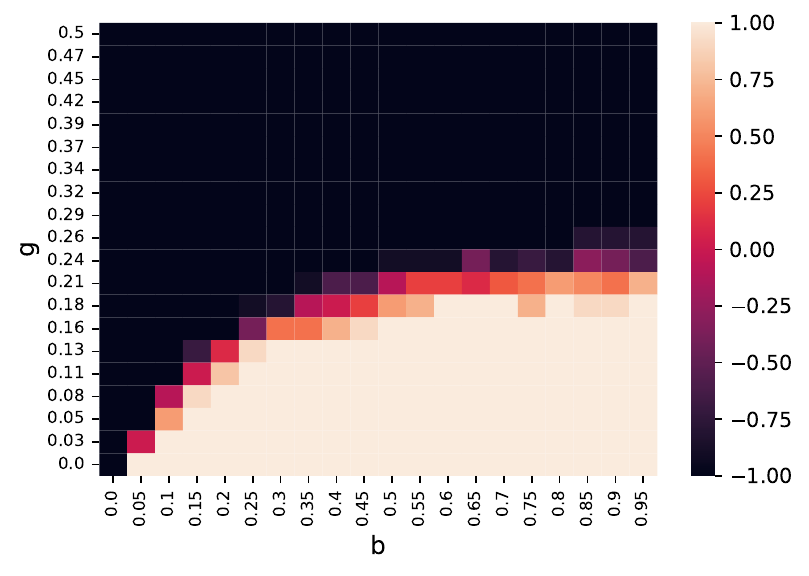}
\includegraphics[width=0.45\linewidth]{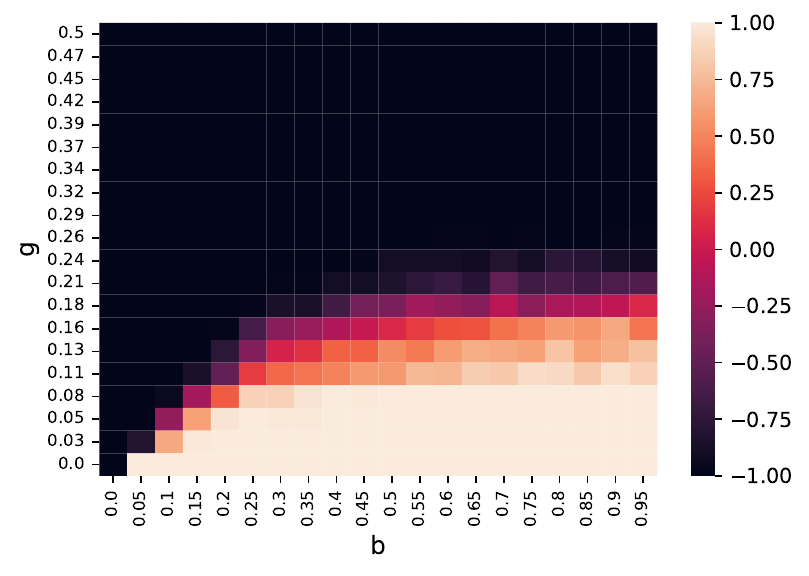}
\caption{The decision boundary $g=G(b)$ by numeric simulation on ER and BA networks. The initial conditions are $r(0)=f(0)=0.01$, $\beta_f=0.1$, $\beta_r=0.05$. The left and right figures are the simulation results in the ER random network and BA scale-free network, respectively. The grid value denotes the simulation results on average based on one hundred experiments. 1 and -1 denote that the intervention is and is not successful in an experiment, respectively.}
\label{ss2}
\end{figure}

The simulation results are shown in Fig.\ref{ss2} for ER and BA models, respectively. We find that the boundary shapes in the case of ER and BA are very similar to the theoretical boundary (Fig. \ref{s3}), but not the same as each other, which means that the network structure also matters a lot in the false information control.

\newpage
\section{Conclusion and discussion}
In this paper, we present an analytical solution for controlling the spreading of false information in various scenarios, using the Susceptible-Infected (SI) framework. The decision boundary for any social network is influenced by various factors, such as the initial number of fake and real agents, their respective spreading rates, and the desired outcome. In addition, in our model, the average degree of the network also plays a crucial role in determining the decision boundary. We also try to assess the intervention costs based on our model, the total intervention cost depends on how we set the intervention cost as a function of the intervention strength $b$. However, only if the intervention cost monopoly increases with the intervention strength $b$, the result should be the sooner we intervene, the lower the overall intervention cost tends to be. Our results demonstrate that the spreading of false information can be controlled in a desired manner based on the theory of mean-field, providing a valuable foundation for realistic policy-making.

However, it is important to note that our results are based on a simplified model that assumes all nodes are homogeneously mixed, thus neglecting the structural complexity of real social networks. In reality, factors such as the network influencers, degree distribution, and correlation between nodes' degrees are vital in determining the spreading of news. Therefore, the accuracy of our results is limited. As demonstrated in the numerical simulation section, the theoretical decision boundary, while similar in shape to the boundaries simulated using ER and BA networks, produces different values. Thus, the theoretical boundary can be regarded as the lower limit of real network boundaries.

% \section{CRediT authorship contribution statement}
% Yu Zhang: Methodology, Formal analysis, Writing – original draft, Conceptualization. Yafei Li: Discussion, Writing – review \& editing; Fanyuan Meng: Editing; Vallarano Nicolò: Review \& Discussion;  Claudio J. Tessone: Methodology.

\section{Declaration of competing interest}
The authors declare that they have no known competing financial
interests or personal relationships that could have appeared to influence the work reported in this paper. 

\section{Data availability}
The datasets used and/or analyzed during the current study are generated and available from the corresponding author upon reasonable request.

\section{Acknowledgements}
Yu Zhang acknowledges the financial support from the China Scholarship Council.
\newpage
\bibliographystyle{unsrt}
\bibliography{reference.bib}

\begin{thebibliography}{10}

\bibitem{heidemann2012online}
Julia Heidemann, Mathias Klier, and Florian Probst.
\newblock Online social networks: A survey of a global phenomenon.
\newblock {\em Computer Networks}, 56(18):3866--3878, 2012.

\bibitem{tambuscio2015fact}
Marcella Tambuscio, Giancarlo Ruffo, Alessandro Flammini, and Filippo Menczer.
\newblock Fact-checking effect on viral hoaxes: A model of misinformation
  spread in social networks.
\newblock In {\em Proceedings of the 24th international conference on World
  Wide Web}, pages 977--982, 2015.

\bibitem{zhaofake}
Z~Zhao et~al.
\newblock Fake news propagates differently from real news even at early stages
  of spreading. epj data sci. 9 (1), 1--14 (2020).

\bibitem{wef}
World~Economic Forum.
\newblock The global risks report, 2023, Retrieved from
  https://www.weforum.org/reports/global-risks-report-2023. Last accessed in
  June, 2023.

\bibitem{shu2020mining}
Kai Shu, Suhang Wang, Dongwon Lee, and Huan Liu.
\newblock Mining disinformation and fake news: Concepts, methods, and recent
  advancements.
\newblock {\em Disinformation, misinformation, and fake news in social media:
  Emerging research challenges and opportunities}, pages 1--19, 2020.

\bibitem{vosoughi2018spread}
Soroush Vosoughi, Deb Roy, and Sinan Aral.
\newblock The spread of true and false news online.
\newblock {\em Science}, 359(6380):1146--1151, 2018.

\bibitem{stahl2006difference}
Bernd~Carsten Stahl.
\newblock On the difference or equality of information, misinformation, and
  disinformation: A critical research perspective.
\newblock {\em Informing Science}, 9:83, 2006.

\bibitem{aimeur2023fake}
Esma A{\"\i}meur, Sabrine Amri, and Gilles Brassard.
\newblock Fake news, disinformation and misinformation in social media: a
  review.
\newblock {\em Social Network Analysis and Mining}, 13(1):30, 2023.

\bibitem{tudjman2003information}
Miroslav Tudjman and Nives Mikelic.
\newblock Information science: Science about information, misinformation and
  disinformation.
\newblock {\em Proceedings of Informing Science+ Information Technology
  Education}, 3:1513--1527, 2003.

\bibitem{petratos2021misinformation}
Pythagoras~N Petratos.
\newblock Misinformation, disinformation, and fake news: Cyber risks to
  business.
\newblock {\em Business Horizons}, 64(6):763--774, 2021.

\bibitem{wardle2017information}
Claire Wardle et~al.
\newblock Information disorder: Toward an interdisciplinary framework for
  research and policy making (2017).
\newblock 2017.

\bibitem{suntwal2020does}
Sandeep Suntwal, Susan Brown, and Mark Patton.
\newblock How does information spread? an exploratory study of true and fake
  news.
\newblock 2020.

\bibitem{bondielli2019survey}
Alessandro Bondielli and Francesco Marcelloni.
\newblock A survey on fake news and rumour detection techniques.
\newblock {\em Information Sciences}, 497:38--55, 2019.

\bibitem{allcott2017social}
Hunt Allcott and Matthew Gentzkow.
\newblock Social media and fake news in the 2016 election.
\newblock {\em Journal of economic perspectives}, 31(2):211--236, 2017.

\bibitem{sharma2019combating}
Karishma Sharma, Feng Qian, He~Jiang, Natali Ruchansky, Ming Zhang, and Yan
  Liu.
\newblock Combating fake news: A survey on identification and mitigation
  techniques.
\newblock {\em ACM Transactions on Intelligent Systems and Technology (TIST)},
  10(3):1--42, 2019.

\bibitem{gelfert2018fake}
Axel Gelfert.
\newblock Fake news: A definition.
\newblock {\em Informal logic}, 38(1):84--117, 2018.

\bibitem{van2022misinformation}
Sander Van Der~Linden.
\newblock Misinformation: susceptibility, spread, and interventions to immunize
  the public.
\newblock {\em Nature Medicine}, 28(3):460--467, 2022.

\bibitem{van2020you}
Sander Van~der Linden, Costas Panagopoulos, and Jon Roozenbeek.
\newblock You are fake news: political bias in perceptions of fake news.
\newblock {\em Media, Culture \& Society}, 42(3):460--470, 2020.

\bibitem{nakov2020can}
Preslav Nakov.
\newblock Can we spot the" fake news" before it was even written?
\newblock {\em arXiv preprint arXiv:2008.04374}, 2020.

\bibitem{mazur9definition}
Viktoria Mazur and Archil Chochia.
\newblock Definition and regulation as an effective measure to fight fake news
  in the european union.
\newblock {\em European Studies}, 9(1):15--40.

\bibitem{guay2022think}
B~Guay, AJ~Berinsky, G~Pennycook, and D~Rand.
\newblock How to think about whether misinformation interventions work.
  psyarxiv, 2022.

\bibitem{he2015modeling}
Zaobo He, Zhipeng Cai, and Xiaoming Wang.
\newblock Modeling propagation dynamics and developing optimized
  countermeasures for rumor spreading in online social networks.
\newblock In {\em 2015 IEEE 35Th international conference on distributed
  computing systems}, pages 205--214. IEEE, 2015.

\bibitem{zhao2011rumor}
Laijun Zhao, Qin Wang, Jingjing Cheng, Yucheng Chen, Jiajia Wang, and Wei
  Huang.
\newblock Rumor spreading model with consideration of forgetting mechanism: A
  case of online blogging livejournal.
\newblock {\em Physica A: Statistical Mechanics and its Applications},
  390(13):2619--2625, 2011.

\bibitem{zhu2020delay}
Linhe Zhu, Wenshan Liu, and Zhengdi Zhang.
\newblock Delay differential equations modeling of rumor propagation in both
  homogeneous and heterogeneous networks with a forced silence function.
\newblock {\em Applied Mathematics and Computation}, 370:124925, 2020.

\bibitem{maleki2021using}
Maryam Maleki, Esther Mead, Mohammad Arani, and Nitin Agarwal.
\newblock Using an epidemiological model to study the spread of misinformation
  during the black lives matter movement.
\newblock {\em arXiv preprint arXiv:2103.12191}, 2021.

\bibitem{acerbi2022research}
Alberto Acerbi, Sacha Altay, and Hugo Mercier.
\newblock Research note: Fighting misinformation or fighting for information?
\newblock 2022.

\bibitem{mahir2019detecting}
Ehesas~Mia Mahir, Saima Akhter, Mohammad~Rezwanul Huq, et~al.
\newblock Detecting fake news using machine learning and deep learning
  algorithms.
\newblock In {\em 2019 7th international conference on smart computing \&
  communications (ICSCC)}, pages 1--5. IEEE, 2019.

\bibitem{qian2018neural}
Feng Qian, Chengyue Gong, Karishma Sharma, and Yan Liu.
\newblock Neural user response generator: Fake news detection with collective
  user intelligence.
\newblock In {\em IJCAI}, volume~18, pages 3834--3840, 2018.

\bibitem{buechel2023misinformation}
Berno Buechel, Stefan Kl{\"o}{\ss}ner, Fanyuan Meng, and Anis Nassar.
\newblock Misinformation due to asymmetric information sharing.
\newblock {\em Journal of Economic Dynamics and Control}, 150:104641, 2023.

\bibitem{bovet2019influence}
Alexandre Bovet and Hern{\'a}n~A Makse.
\newblock Influence of fake news in twitter during the 2016 us presidential
  election.
\newblock {\em Nature communications}, 10(1):7, 2019.

\bibitem{gilligan2008sustainable}
Christopher~A Gilligan.
\newblock Sustainable agriculture and plant diseases: an epidemiological
  perspective.
\newblock {\em Philosophical Transactions of the Royal Society B: Biological
  Sciences}, 363(1492):741--759, 2008.

\bibitem{chaudhury2012spread}
Arpan Chaudhury, Partha Basuchowdhuri, and Subhashis Majumder.
\newblock Spread of information in a social network using influential nodes.
\newblock In {\em Advances in Knowledge Discovery and Data Mining: 16th
  Pacific-Asia Conference, PAKDD 2012, Kuala Lumpur, Malaysia, May 29--June 1,
  2012, Proceedings, Part II 16}, pages 121--132. Springer, 2012.

\bibitem{alassad2019finding}
Mustafa Alassad, Muhammad~Nihal Hussain, and Nitin Agarwal.
\newblock Finding fake news key spreaders in complex social networks by using
  bi-level decomposition optimization method.
\newblock In {\em Modeling and Simulation of Social-Behavioral Phenomena in
  Creative Societies: First International EURO Mini Conference, MSBC 2019,
  Vilnius, Lithuania, September 18--20, 2019, Proceedings 1}, pages 41--54.
  Springer, 2019.

\bibitem{shu2019studying}
Kai Shu, H~Russell Bernard, and Huan Liu.
\newblock Studying fake news via network analysis: detection and mitigation.
\newblock {\em Emerging research challenges and opportunities in computational
  social network analysis and mining}, pages 43--65, 2019.

\bibitem{zhou2019network}
Xinyi Zhou and Reza Zafarani.
\newblock Network-based fake news detection: A pattern-driven approach.
\newblock {\em ACM SIGKDD explorations newsletter}, 21(2):48--60, 2019.

\bibitem{ruchansky2017csi}
Natali Ruchansky, Sungyong Seo, and Yan Liu.
\newblock Csi: A hybrid deep model for fake news detection.
\newblock In {\em Proceedings of the 2017 ACM on Conference on Information and
  Knowledge Management}, pages 797--806, 2017.

\bibitem{budak2011limiting}
Ceren Budak, Divyakant Agrawal, and Amr El~Abbadi.
\newblock Limiting the spread of misinformation in social networks.
\newblock In {\em Proceedings of the 20th international conference on World
  wide web}, pages 665--674, 2011.

\bibitem{traberg2022psychological}
Cecilie~S Traberg, Jon Roozenbeek, and Sander van~der Linden.
\newblock Psychological inoculation against misinformation: Current evidence
  and future directions.
\newblock {\em The ANNALS of the American Academy of Political and Social
  Science}, 700(1):136--151, 2022.

\bibitem{roozenbeek2019fake}
Jon Roozenbeek and Sander Van Der~Linden.
\newblock The fake news game: actively inoculating against the risk of
  misinformation.
\newblock {\em Journal of risk research}, 22(5):570--580, 2019.

\bibitem{lewandowsky2021countering}
Stephan Lewandowsky and Sander Van Der~Linden.
\newblock Countering misinformation and fake news through inoculation and
  prebunking.
\newblock {\em European Review of Social Psychology}, 32(2):348--384, 2021.

\bibitem{daley1964epidemics}
Daryl~J Daley and David~G Kendall.
\newblock Epidemics and rumours.
\newblock {\em Nature}, 204:1118--1118, 1964.

\bibitem{maki1973mathematical}
Daniel~P Maki and Maynard Thompson.
\newblock Mathematical models and applications: with emphasis on the social
  life, and management sciences.
\newblock Technical report, 1973.

\bibitem{wang2013rumor}
Jiajia Wang, Laijun Zhao, Rongbing Huang, and Yucheng Chen.
\newblock Rumor spreading model on social networks with consideration of
  remembering mechanism.
\newblock In {\em IET International Conference on Smart and Sustainable City
  2013 (ICSSC 2013)}, pages 27--31. IET, 2013.

\bibitem{liao2014can}
Q~Vera Liao and Wai-Tat Fu.
\newblock Can you hear me now? mitigating the echo chamber effect by source
  position indicators.
\newblock In {\em Proceedings of the 17th ACM conference on Computer supported
  cooperative work \& social computing}, pages 184--196, 2014.

\bibitem{jin2013epidemiological}
Fang Jin, Edward Dougherty, Parang Saraf, Yang Cao, and Naren Ramakrishnan.
\newblock Epidemiological modeling of news and rumors on twitter.
\newblock In {\em Proceedings of the 7th workshop on social network mining and
  analysis}, pages 1--9, 2013.

\bibitem{fan2020novel}
Dongmei Fan, Guo-Ping Jiang, Yu-Rong Song, and Yin-Wei Li.
\newblock Novel fake news spreading model with similarity on pso-based
  networks.
\newblock {\em Physica A: Statistical Mechanics and its Applications},
  549:124319, 2020.

\bibitem{de2015exploiting}
Giuseppe De~Martino and Serena Spina.
\newblock Exploiting the time-dynamics of news diffusion on the internet
  through a generalized susceptible--infected model.
\newblock {\em Physica A: Statistical Mechanics and its Applications},
  438:634--644, 2015.

\bibitem{abdullah2011epidemic}
Saeed Abdullah and Xindong Wu.
\newblock An epidemic model for news spreading on twitter.
\newblock In {\em 2011 IEEE 23rd International Conference on Tools with
  Artificial Intelligence}, pages 163--169. IEEE, 2011.

\bibitem{rodrigues2015viral}
Helena~Sofia Rodrigues and Manuel~Jos{\'e} Fonseca.
\newblock Viral marketing as epidemiological model.
\newblock {\em arXiv preprint arXiv:1507.06986}, 2015.

\bibitem{rossman2021network}
Gabriel Rossman and Jacob~C Fisher.
\newblock Network hubs cease to be influential in the presence of low levels of
  advertising.
\newblock {\em Proceedings of the National Academy of Sciences},
  118(7):e2013391118, 2021.

\end{thebibliography}
% \nocite{*}

\end{document}